# Unconventional charge-to-spin conversions in graphene/MoTe$_2$ van der Waals heterostructures

Nerea Ontoso[1], C.K. Safeer[1], Franz Herling[1], Josep Ingla-Aynés[1], Haozhe Yang[1], Zhendong Chi[1], Iñigo Robredo[2], Maia G. Vergniory[2,3], Fernando de Juan[3,4], M. Reyes Calvo[5,6,*], Luis E. Hueso[1,4], Fèlix Casanova[1,4,*]

[1] CIC nanoGUNE BRTA, 20018 Donostia-San Sebastian, Basque Country, Spain.
[2] Max Planck Institute for Chemical Physics of Solids, Dresden D-01187, Germany
[3] Donostia International Physics Center, 20018 Donostia-San Sebastian, Basque Country, Spain
[4] IKERBASQUE, Basque Foundation for Science, 48009 Bilbao, Basque Country, Spain.
[5] Departamento de Física Aplicada, Universidad de Alicante, 03690 Alicante, Spain.
[6] Instituto Universitario de Materiales de Alicante (IUMA), Universidad de Alicante, 03690 Alicante, Spain.
*E-mail reyes.calvo@ua.es; f.casanova@nanogune.eu

## Abstract

Spin-charge interconversion (SCI) is a central phenomenon to the development of spintronic devices from materials with strong spin-orbit coupling (SOC). In the case of materials with high crystal symmetry, the only allowed SCI processes are those where the spin current, charge current and spin polarization directions are orthogonal to each other. Consequently, standard SCI experiments are designed to maximize the signals arising from the SCI processes with conventional mutually orthogonal geometry. However, in low-symmetry materials, certain non-orthogonal SCI processes are also allowed. Since the standard SCI experiment is limited to charge current flowing only in one direction in the SOC material, certain allowed SCI configurations remain unexplored. In this work, we performed a thorough SCI study in a graphene-based lateral spin valve combined with low-symmetry MoTe$_2$. Due to a very low contact resistance between the two materials, we could detect SCI signals using both a standard configuration, where the charge current is applied along the MoTe$_2$, and a recently introduced (3D-current) configuration, where the charge current flow can be controlled in three directions within the heterostructure. As a result, we observed three different SCI components, one orthogonal and two non-orthogonal, giving new insight into the SCI processes in low-symmetry materials. The large SCI signals obtained at room temperature, along with the versatility of the 3D-current configuration, provide feasibility and flexibility to the design of the next generation of spin-based devices.

## I. INTRODUCTION

Since its momentous discovery in 2004, graphene has become a unique platform to investigate new physical phenomena[1,2]. Its isolation opened the door to the discovery and study of a huge family of two-dimensional (2D) materials that can host a plethora of properties[3–10]. In the field of spintronics, graphene has proven to be an ideal candidate to transport spin currents over long distances, due to its low spin-orbit coupling (SOC)[11–14], which in turn limits its capability for the manipulation of spin currents. However, SOC can be enhanced in graphene by proximity to another 2D material in van der Waals heterostructures,[15–20] leading to weak antilocalization,[21–25] spin lifetime anisotropy,[26–28] and enabling the electrical control of spin currents[29,30] and of the SOC-induced spin precession[31]. Spin-orbit proximity in graphene also causes the spin Hall[32–35] (SHE) and

Edelstein[34–39] (EE) effects, giving rise to electrically controllable spin-charge interconversion (SCI). This is a crucial ingredient to achieve magnetic-field-free switching[40] and readout[41,42] of magnetic elements in memory and logic devices.

The spin Hall conductivity tensor $\sigma$ relates the electric field ($E$), associated to the applied charge current, with the generated spin current density by $j_i^k = \sigma_{ij}^k E_j$, where $i$ corresponds to the spin current direction, $j$ to the charge current direction and $k$ to the spin polarization direction. The non-zero components of the tensor determine the allowed geometrical configurations of the SCI due to the SHE. In materials with high-crystal symmetry, where at least two mirror planes are present, the only non-zero elements of $\sigma_{ij}^k$ fulfill that $i \neq j \neq k$, i.e., charge current, spin current and spin polarization directions are mutually orthogonal. However, in low-symmetry materials, additional non-orthogonal components are allowed[43–46]. That is the case for the transition metal dichalcogenide MoTe$_2$ in the 1T' crystal phase, which is characterized by only one mirror plane ($\mathcal{M}_a$), perpendicular to the Mo-chain, which lies along the *y*-direction in the experimental configuration, and a screw axis parallel to it[47]. In this case, when the spin polarization is in plane and perpendicular to the mirror plane ($k$ along *y*-direction), the $\sigma_{ij}^k$ components for which the spin current is parallel to the applied charge current ($i = j$) are also allowed giving $\sigma_{ii}^y \neq 0$ (see Supplemental Material section S1 and Ref. [46]). In addition to SHE, the EE can also lead to SCI. In SOC materials with broken inversion symmetry, the flow of charge current is expected to generate spin accumulation in the system. In a graphene-based van der Waals heterostructure, this effect can also occur in the proximitized graphene [34–37,39]. The symmetries that govern the permitted SCI processes via EE depend on the mirror alignment between the graphene and the SOC material[48–53]. In twisted heterostructures, besides the conventional EE where the charge current induces a perpendicular spin polarization, the broken mirror symmetry leads to an unconventional EE with parallel spin polarization. Experimentally, such unconventional EE has been recently observed in graphene/MoTe$_2$ (Ref. [54]), graphene/WTe$_2$ (Ref. [55]), and graphene/NbSe$_2$ (Ref. [56]) heterostructures.

So far, 1T'-MoTe$_2$ has been studied as a SCI material using non-local lateral spin valve (LSV)[39,54] and spin-orbit torque experiments[57,58]. While most artifacts induced by the local charge current are avoided in non-local LSVs, the injection of the charge current in the SOC material is limited to one direction, making certain SCI configurations to remain unexplored. For SCI experiments using non-local LSVs, based on spin absorption phenomenon, a versatile measurement geometry is required to study unexplored SCI components. Recently, a new measurement configuration, where the charge current propagates in three dimensions (3D) was implemented, leading to the observation of different SCI components in graphene/WTe$_2$ (Ref. [59]) and graphene/BiTeBr (Ref. [60]) heterostructures. In these studies, the charge current was applied between the SOC material and the graphene channel (Fig. 3a and 3c). These experiments, however, only returned a measurable signal when the interface resistance was high enough and the spin absorption into the SOC material was suppressed, implying that the simultaneous detection of SCI in the conventional and the new 3D-current configuration could not be achieved.

In this work, we combine the conventional configuration with the recently introduced 3D-current configuration to measure and characterize SCI in a graphene/MoTe$_2$ van der Waals heterostructure. The low interface resistance between graphene and MoTe$_2$ allows us to measure SCI in both configurations in the same device up to room temperature. We discover that spin generation occurs when charge current flows through the MoTe$_2$ flake in any direction, gaining deeper insights into the multi-directional SCI processes in graphene/MoTe$_2$. Whereas the SCI component due to a non-orthogonal arrangement of spin polarization and charge current has been measured before,[54] new components due to both orthogonal and non-orthogonal SCI geometries are measured here. Specifically, we detect SCI signals corresponding to the cases where a charge current propagating along the mirror plane (*b*-crystal axis in Fig.1a, which corresponds to the *x*-direction in our space coordinates) or along the vertical stacking (*z*-direction) generates a spin current with the same in-plane orthogonal polarization (*a*-crystal axis in Fig.1a, corresponding to the *y*-direction). The origin of the former contribution is compatible with SHE in MoTe$_2$ or EE in the proximitized graphene, while the SHE in MoTe$_2$ is the only allowed mechanism that explains the unexplored $\sigma_{zz}^y$ SHE component in MoTe$_2$, additionally confirmed using DFT calculations. The large SCI signals measured here at room temperature, together with the versatility in the 3D-current configurations, makes MoTe$_2$/graphene heterostructures ideal candidates to be integrated in functional 2D spintronic devices.

## II. EXPERIMENTAL DETAILS

The LSV device used for our measurement is shown in Fig. 1b. The device is similar to those used in a related work by some of the authors (Ref. 54). At first, an elongated MoTe$_2$ flake of ~20 nm thickness - MoTe$_2$ tends to cleave along the Mo-chain[54] (pink zig-zag line along the *a*-crystal axis in Fig. 1a)- was mechanically exfoliated in an inert atmosphere and stamped on top of an exfoliated few-layer graphene flake using the dry-viscoelastic techinque[61]. In this way, MoTe$_2$ is not exposed to air, preventing oxidation, and keeping a clean interface between MoTe$_2$ and graphene. The width of the MoTe$_2$ flake on top of the graphene channel is ~1 μm. After the stamping, metallic Ti(5 nm)/Au(100 nm) electrodes (labeled with letters in Fig. 1b) contacting MoTe$_2$ and graphene were fabricated via e-beam lithography, thermal evaporation, and lift-off. Subsequently, ferromagnetic (FM) electrodes (labeled with numbers in Fig. 1b) -TiO$_x$(0.3 nm)/Co(35 nm)- were fabricated on top of graphene , using the same technique. TiO$_x$ was prepared by air exposure of a Ti layer, as described elsewhere[54]. The width of TiO$_x$/Co electrodes is approximately 300 nm and 400 nm on the left and right sides of the MoTe$_2$ flake, respectively.

To characterize our MoTe$_2$ flake, we have measured its resistance (*R*) by applying a current between contacts A and B ($I_{AB}$) and measuring the voltage between the same electrodes ($V_{AB}$). We label this geometry as $V_{AB}I_{AB}$. As shown in Fig. 1c, the resistance of the MoTe$_2$ flake ($R_{AB} = V_{AB}/I_{AB}$) has a temperature dependence compatible with a semimetallic behavior,[62] . Resistance curves for cooling and heating sweeps perfectly overlap, showing no fingerprint of the structural transition from 1T' to 1T$_d$ phase[57,63] otherwise reported for the bulk material[64] (see Supplemental Material section S2 for further characterization with polarized Raman spectroscopy). The graphene flake resistance, measured in a 4-point configuration ($V_{21}I_{CD}$), is also plotted in Fig. 1c. The quality of the interface can be inferred by measuring the interface resistance between graphene and MoTe$_2$. By applying a charge current between one side of MoTe$_2$ and graphene ($I_{BD}$) and measuring the voltage between the other sides ($V_{AC}$), the interface resistance ($R_{int} = V_{AC}/I_{BD}$) is probed. As shown in Fig. 1c, $R_{int}$ shows very low values, even negative at room temperature and increase up to 100 Ω at 10 K, meaning that the interface resistance is smaller than the graphene square resistance [54].

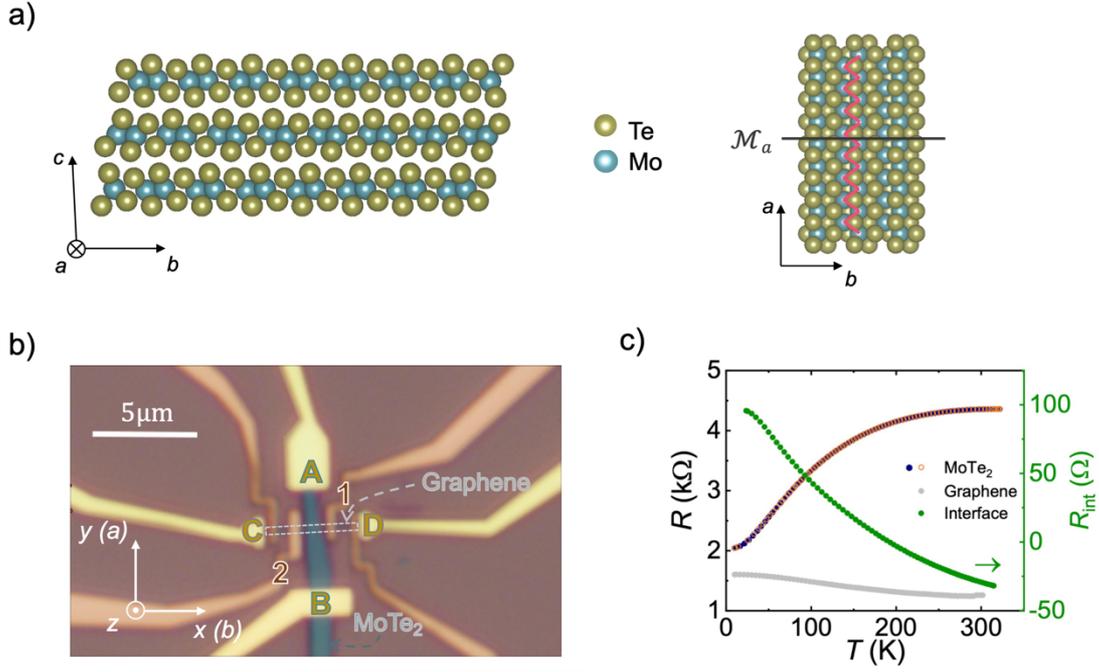

**Fig. 1.** (a) Sketch of the MoTe$_2$ crystal structure in the 1T' phase. This phase is characterized by a tilt of 93.83° in the $c$-crystal axis and a Mo-chain along the $a$-crystal axis (pink zig-zag line). The $a$-axis corresponds to the preferential cleaving direction during mechanical exfoliation, giving rise to elongated flakes suitable for the LSV devices. The 1T' phase hosts a mirror in the $b$-$c$ plane ($\mathcal{M}_a$) and a screw axis along the $b$-crystal axis (not represented). (b) Optical-microscope image of the device. The barely visible narrow graphene flake along $x$, represented by the dashed rectangle, is under the MoTe$_2$ flake (blue). The Ti/Au electrodes are labeled with letters and the FM TiO$_x$/Co electrodes with numbers. The relation between the crystal axis and the space coordinates is as follows: $b = x$ and $a = y$. It should be noted that the $c$ crystal axis is not along the $z$-direction. (c) Two-point resistance as a function of temperature of the MoTe$_2$ flake when cooling down (orange open circles) and warming up (blue filled circles), measured in the $V_{AB}/I_{AB}$ configuration. The 4-point resistance of the few-layer graphene flake as a function of temperature, measured in the $V_{21}/I_{CD}$ configuration, is also shown (grey circles). The green circles correspond to the interface resistance between graphene and MoTe$_2$, measured in the $V_{AC}/I_{BD}$ configuration.

## III. RESULTS AND DISCUSSION

CI has been observed in graphene due to spin-orbit proximity effect. Most of these reported experiments uses a LSV based on graphene and a semiconducting high-SOC van der Waals material, in which graphene is shaped into a Hall bar and contacted to probe SCI voltage from the proximitized region [32–39]. However, if graphene is combined with a conducting (metallic or semimetallic) van der Waals material, SCI can be measured by directly probing the voltage across both materials. In this case, disentangling the origin of each SCI process remains challenging as it can either occur via EE or SHE in the proximitized graphene[36,39,55,56,65] or in the high-SOC material itself via the SHE[38,54,66,67]. In the standard non-local SCI measurement configuration[54,68,69], the charge current ($I_c$) is applied along the SOC material ($y$-direction). In highly symmetric materials, $I_c$ generates a spin current via the SHE in the vertical direction ($z$-direction) with spin polarization along the $x$-direction (corresponding to the spin Hall conductivity tensor component $\sigma_{zy}^x$). The spin current flows into the graphene channel and is detected with a FM electrode as a non-local voltage ($V_{NL}$). To detect the spin current, the magnetization of the FM has to be pulled in the same direction as the spin polarization and, thus, the magnetic field is applied along the in-plane hard axis ($x$-direction). When the SOC material has lower crystal symmetry, spin polarization in other directions could also be generated. In the 1T' crystal phase, non-orthogonal components are allowed with spins polarized along the Mo chain (*y*-direction, see Supplemental Material.). Hence, for this material, it is also of interest to keep the magnetization of the FM along its easy axis (*y*-direction) while the charge current is applied along different directions within the high-SOC material.

### A. Standard configuration

We first study SCI by applying the charge current along the MoTe$_2$ flake [i.e., along the *y*-direction, Fig. 2a] while sweeping the magnetic field along the FM easy axis ($H_y$). The non-local resistance ($R_{NL}=V_{NL}/I_c$) dependence with $H_y$ shows a hysteresis loop with two clear jumps at the switching field of the FM when using contact 2 (on the left side of the MoTe$_2$ flake) as a detector (Fig. 2b). The amplitude of the signal is given by $\Delta R_{NL}$, which in this case can only arise from the charge current $I_c$ that flows along the −*y*-direction, denoted as $I_c^{-y}$, either along the MoTe$_2$ flake or along the proximitized graphene. When $H_y$ is positive (negative), $R_{NL}$ takes its lower (higher) value. Hence, the injected spins are polarized along the -*y* direction. Note that we have assumed that the graphene/FM contact spin polarization is positive. We keep this convention through all the manuscript. Therefore, we can conclude that a *y*-spin polarization is generated by applying $I_c$ along the same direction in the MoTe$_2$ flake.

Three possible mechanisms could be behind this spin generation: (*i*) EE in the proximitized graphene, (*ii*) SHE in MoTe$_2$ with spin current along the *z*-direction, or (*iii*) SHE in MoTe$_2$ with spin current along the *x*-direction. Concerning case (*i*), a spin accumulation polarized along -*y* must be generated by a perpendicular charge current (*x*- or *z*-direction) in the conventional EE. However, in this case, the charge current and the spin polarization are parallel, which is forbidden by the crystal symmetry of the graphene. If we consider cases (*ii*) and (*iii*) both SHE components ($\sigma_{xy}^y$ or $\sigma_{zy}^y$ respectively) should be zero, since they are forbidden by the crystal symmetries of MoTe$_2$ (see Supplemental Material).

To further understand the possible mechanism, $R_{NL}$ is measured using contact 1 as a detector (on the right side of MoTe$_2$ flake) while the magnetic field is swept along the FM easy axis as represented in Fig. 2c, a configuration which probes the spin current flowing in the opposite direction than using contact 2. $\Delta R_{NL}$ shows the same sign as the one detected on the left side. The absence of a sign reversal further proves that the spin current along the *x*-direction does not play any role in the SCI. Accordingly, two mechanisms remain as possible origins of the signal: an unconventional SHE in MoTe$_2$ with spin-current along the *z*-direction ($\sigma_{zy}^y$) or an unconventional EE at the proximitized graphene (that does not depend on spin current directions), where the spin polarization and charge current are parallel. Both scenarios are incompatible with the processes allowed by the crystal symmetry of both MoTe$_2$ and graphene.

Therefore, the symmetry of the system must be broken to allow such unconventional SCI. A possible origin for symmetry breaking is shear strain, which may develop during the fabrication process, as suggested in Ref. [57]. However, this signal has now been reproduced in several graphene/MoTe$_2$ samples (this work and Ref. [54]) and even in different graphene/TMD heterostructures[55,56]. Shear strain is expected to be sample-dependent,[70] making this hypothesis unlikely. Another possible mechanism is that the misalignment between the mirror planes of graphene and MoTe$_2$ creates a non-symmetric interface. This enables unconventional EE in the proximitized graphene, for which the spin polarization is parallel to the charge current direction [48–53].

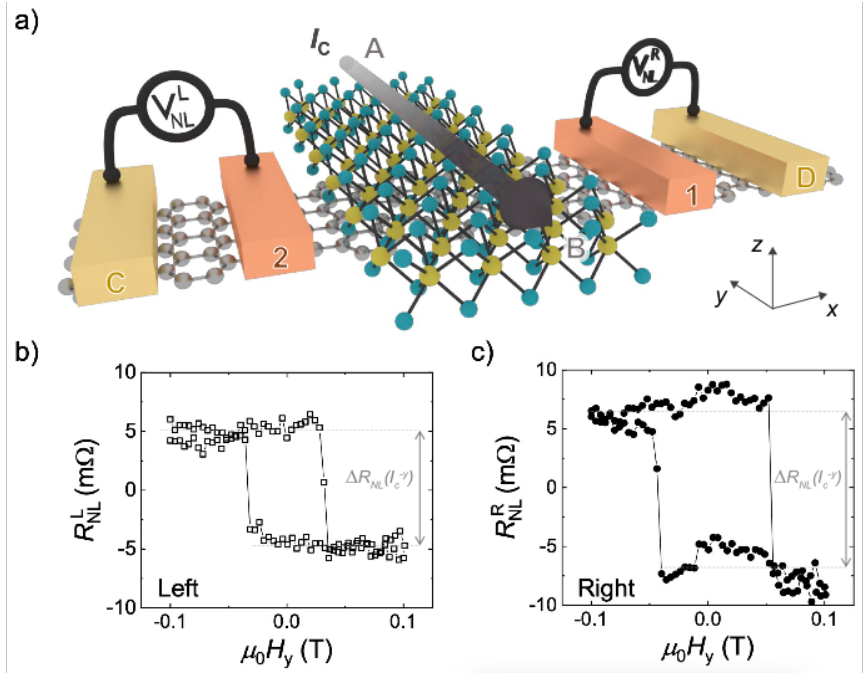

**Fig. 2. (a)** Sketch of the standard SCI measurement configurations that can be performed in the heterostructure. The charge current ($I_C$) is applied along $MoTe_2$ and the voltage is probed using the FM electrode on the left or on the right sides of the $MoTe_2$ flake. **(b)** Non-local resistance ($R_{NL}$) measured using the left Co contact 2 as a detector ($V_{NL}^L$) while sweeping the magnetic field along the in-plane easy axis ($y$-direction). **(c)** The similar measurement using the right Co electrode 1 as detector ($V_{NL}^R$). Baselines of 110 mΩ and 26 mΩ are removed, respectively. All the measurements are performed at 300 K.

### B. 3D-current configuration

So far, we have only explored SCI arising from a current flowing along the $-y$-direction ($I_c^{-y}$). To further access other SCI processes, the current path in $MoTe_2$ must include components along the three different space coordinates. We implement a protocol [38,57,58,65] (which we will call 3D-current configuration) where $I_c$ is injected from $MoTe_2$ into graphene through their interface using one Ti/Au electrode on $MoTe_2$ (A or B) and other one on graphene (C or D). In this configuration, $I_c$ propagates along $x$, $y$ and $z$, and spin currents arising from current flowing in all three directions can be detected. To separate their contributions to spin polarization, we test different non-local measurement configurations by reversing each $x$, $y$ or $z$ component of $I_c$ independently. When one of the components of $I_c$ is reversed, the spin polarization and the associated non-local voltage in the detector resulting from the charge current must change sign, regardless of the SCI origin.

The four possible charge current paths within the graphene/$MoTe_2$ heterostructure when contact 1 is used as the FM detector are sketched in Fig. 3a. The measured $R_{NL}$ as a function of $H_y$ for each of those current paths is plotted in Fig. 3b, where hysteresis loops are represented with the same color as the respective current path. In this specific measurement configuration, $R_{NL}$ signals are proportional to the number of $y$-polarized spins arriving to the detector. In other words, our experimental configuration detects SCI processes which generate $y$-polarized spins, which may arise from any charge current propagation direction. To distinguish the charge current direction creating the spin polarization along $y$, each charge current path can be spatially decomposed as follows: dark red path corresponds to $I_c(-x, -y, -z)$, light red to $I_c(+x, +y, +z)$, dark blue to $I_c(-x, +y, -z)$ and light blue to $I_c(+x, -y, +z)$. As shown in Fig. 3b, dark red and dark blue curves show the same hysteresis loop sign, while light red and light blue curves also show the same sign but opposite to the dark curves. Considering the correspondence between signs of the loops and the charge current directions, we can conclude that the sign change of the $R_{NL}$ signal may arise from a reversal of the charge current either along the *x*- or the *z*-direction.

To shed further light on the SCI origin, another set of measurements with the 3D-current configuration is implemented using contact 2 as the FM detector, which allows us to access four extra current paths (see Fig. 3c). Figure 3d shows $R_{NL}$ as a function of $H_y$ for the four extra configurations. In this case, the decomposition of the charge current is as follows: dark red path corresponds to $I_c$ (+*x*, −*y*, −*z*), light red to $I_c$ (−*x*, +*y*, +*z*), dark blue to $I_c$ (+*x*, +*y*, −*z*) and light blue to $I_c$ (−*x*, −*y*, +*z*). If we apply the same sign comparison protocol used for measurements shown in Fig. 3b, we can again conclude that the $R_{NL}$ signal originates from the charge current either along the *x*- or the *z*-direction. To distinguish between these two possible origins, we need to further compare the two sets of measurements shown in Fig. 3b and 3d. Table 1 contains a summary of the different measurement configurations, the respective spatial components of $I_c$, and the amplitude of the resulting non-local signal loop ($\Delta R_{NL}$) at the detector. It is important to notice that only the $I_c$ component along the *x*-direction is reversed when comparing the two sets of data obtained from detector 1 and 2. In other words, in the configurations represented with the same color in Fig. 3a and Fig. 3c respectively the charge current follows the same path except for a reversal in the x direction. If their corresponding same-color $R_{NL}$ signals in Fig. 3b and Fig. 3d are compared, a reverse in the $R_{NL}$ signal is observed for each of the four non-local resistance configurations. This implies that it is the charge current along the *x*-direction ($I_c^{\pm x}$) that dominates the generation of spin polarization along the ∓*y*-direction. Hanle spin precession measurements corroborate the spin origin of the non-local signals detected both for the standard and the 3D-current measurement configurations (see Supplemental Material section S3).

### C. Analysis of the signals and discussion of the mechanisms

The possible mechanisms behind SCI with charge current along *x* and spin polarization along *y* are SHE in MoTe$_2$ or EE in the proximitized graphene. For the former, a charge current along the *x*-direction will generate a spin current along the *z*-direction with spin polarization along *y* (corresponding to $\sigma_{zx}^y$). For the latter, a charge current along *x* generates a spin accumulation with spin polarization along *y*. In contrast to the signal measured using the standard SCI configuration (Fig. 2), and regardless of the SHE or EE origin, this signal is allowed by symmetry (see Supplemental Material.) and has not been detected before.

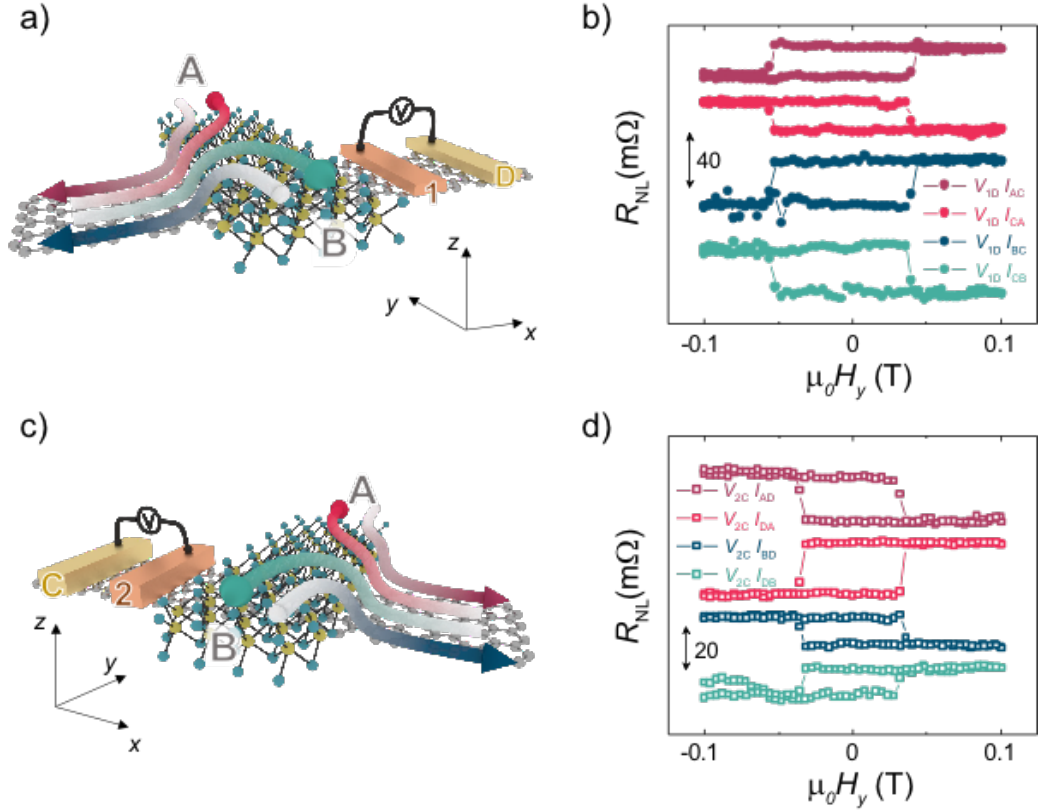

**Fig. 3. (a)** Sketch of all 3D-current SCI measurement configurations using contact 1 as a FM detector. The arrows across the MoTe$_2$/graphene represent the four different current paths. **(b)** Non-local resistance measured using contact 1 as a FM detector while sweeping the magnetic field along the FM easy axis ($H_y$) for the different current paths. The curves have been vertically shifted for clarity. The arrow represents an amplitude of 40 mΩ. **(c)** Sketch of all 3D-current SCI measurement configurations using contact 2 as a FM detector. **(d)** Non-local resistance measured using contact 2 as FM while sweeping the magnetic field along the FM easy axis ($H_y$) for the different current paths. The arrow represents an amplitude of 20 mΩ. All the measurements are performed at 300 K.

Using both the standard (Fig. 2a and 2b) and the 3D-current (Fig. 3a and 3c) SCI measurement configurations in the graphene/MoTe$_2$ heterostructure, we have observed how both x- and y-charge current directions ($I_c^{\pm x}$ and $I_c^{\pm y}$ respectively) induce spins polarized along the y-direction. To quantitatively compare them, the amplitude of the SCI signal ($\Delta R_{NL}$), which is defined as the difference between the mean value of the non-local resistance at the two saturating states, is extracted for each configuration and shown in Table 1. For each FM detector, two configurations result in positive $\Delta R_{NL}$ while the other two show opposite sign. It is remarkable that the absolute value of the amplitude for the signals with the same sign are not the same. The difference in $\Delta R_{NL}$ between the two signals with positive (negative) amplitudes, that is approximately 10 mΩ in both cases, must arise from the SCI of the charge current along the y-direction as discussed next.

From the standard SCI measurement configuration (Fig. 2), we know that there is a SCI signal with spin polarization along y when the charge current is applied along the $-y$-direction ($I_c^{-y}$) and $\Delta R_{NL}$ of this contribution is negative, being positive when the current flows in the $+y$-direction ($I_c^{+y}$). In the 3D current experiments, the contribution to SCI is measured together with the signal coming from the charge current flowing in the x-direction. If we compare the dark blue and dark red configurations in Fig. 3a or Fig. 3c, the only difference in the charge current path between them is the direction of the charge current along the *y*-direction. For example, in the case of contact 1 working as a detector (Fig. 3a-b), the dark blue configuration corresponds to $I_c$ (−*x*, +*y*, −*z*) while the dark red case corresponds to $I_c$ (−*x*, −*y*, −*z*). Since $\Delta R_{NL}$ is positive, we conclude that a charge current along the −*x*-direction ($I_c^{-x}$) will generate a spin polarization along the +*y* axis. In contrast, a charge current along the −*y*-direction ($I_c^{-y}$) will generate a spin polarization along the −*y*-direction and $I_c^{+y}$ will generate spin polarization along +*y*. A similar scenario can be applied to explain the difference between the two negatives $\Delta R_{NL}$ for the light colored curves. In other words, for each pair of dark or light curves measured in detector 1, the contributions to $\Delta R_{NL}$ arising from $I_c^{\pm x}$ and $I_c^{\pm y}$ have the same sign for blue curves, therefore they add up. In contrast, for the red curves the contributions to $\Delta R_{NL}$ arising from $I_c^{\pm x}$ and $I_c^{\pm y}$ present opposite sign, subtracting their values (Fig. 4). By considering the contribution to the amplitude of each hysteresis loop coming from $I_c^{\pm x}$ and $I_c^{\pm y}$, that is $\Delta R_{NL}(I_c^{\pm x})$ and $\Delta R_{NL}(I_c^{\pm y})$ respectively, the ratio of the absolute values of the SCI signal amplitudes $\Delta R_{NL}(I_c^{\pm x})/\Delta R_{NL}(I_c^{\pm y})$ can be obtained. A similar analysis is also shown in Fig. 4 for the measurements using FM detector 2 explained in Fig. 3c and 3d.

| Configuration | Detector | $I_c$ | $\Delta R_{NL}$ (mΩ) |
|---|---|---|---|
| Standard configuration | Contact 1 | ● (0, −*y*, 0) | −13.6 |
| | Contact 2 | ☐ (0, −*y*, 0) | −9.1 |
| 3D-current configuration | Contact 1 | ● (−*x*, −*y*, −*z*) | 22.2 |
| | | ● (+*x*, +*y*, +*z*) | −21.8 |
| | | ● (−*x*, +*y*, −*z*) | 32.4 |
| | | ● (+*x*, −*y*, +*z*) | −30.7 |
| | Contact 2 | ☐ (+*x*, −*y*, −*z*) | −21.4 |
| | | ☐ (−*x*, +*y*, +*z*) | 24.1 |
| | | ☐ (+*x*, +*y*, −*z*) | −13.1 |
| | | ☐ (−*x*, −*y*, +*z*) | 10.6 |

**Table 1.** Summary of the different measurement configurations carried out in the graphene/MoTe$_2$ heterostructure. For each configuration, the components of $I_c$ with the corresponding symbol in the plot, and the respective amplitude of the signals are detailed.

On another comparison, the amplitude of the signal measured using contact 1 as detector is approximately 1.5 times larger than if detected using contact 2 (see Table 1). This difference can be attributed to the different degree of polarization of the two FM electrodes, or to the different distance between the detector and the MoTe$_2$ flake. However, in the 3D configuration, we observe also large differences between signals measured with the left and the right detector which cannot be solely attributed to the different degree of polarization or the different distance to the Co electrodes. For example, if we consider that only x and y-flowing currents contribute to spin generation, we can calculate a ratio $\Delta R_{NL}(I_c^{\pm x})/\Delta R_{NL}(I_c^{\pm y})$, which is significantly smaller for detector 2 than the one obtained for detector 1. If the difference arose just from the polarization or the distance to the FM leads, the ratio $\Delta R_{NL}(I_c^{\pm x})/\Delta R_{NL}(I_c^{\pm y})$ should be the same for both detection schemes. Interestingly, and even if we cannot fully discard contributions from geometrical factors or from inhomogeneities in the spin absorption of the MoTe$_2$-graphene interface, the pattern on the differences between detector 1 and detector 2 for the signals can also be explained by considering the presence of SCI from $I_c^{\pm z}$. If we now consider that the amplitude of each hysteresis loop comes from the presence of SCI of the spins polarized along *y* for three charge current directions, SCI signal coming from $I_c^{\pm z}$, can be extracted, which is comparable to the one of $\Delta R_{NL}(I_c^{\pm y})$, being both smaller than $\Delta R_{NL}(I_c^{\pm x})$. $\Delta R_{NL}(I_c^{+z})$ and $\Delta R_{NL}(I_c^{+x})$ show the same sign, opposite to the one of $\Delta R_{NL}(I_c^{+y})$ (Table 1 and Fig. 4). In this scenario, when using contact 1 as detector, the contributions of $\Delta R_{NL}(I_c^{\pm x})$ and $\Delta R_{NL}(I_c^{\pm z})$ present the same sign for each situation, enlarging the total amplitude of the hysteresis loop. However, in the case of contact 2, both contributions present opposite sign, reducing the total amplitude of the SCI signal. The origin of this SCI component arising from $I_c^{\pm z}$ can only be attributed to SHE in MoTe$_2$ where the charge current is along the *z*-direction, parallel to the spin current direction, and the injected spins are polarized along the *y*-direction. This configuration corresponds to the $\sigma_{zz}^y$ tensor element, which is allowed by the crystal symmetry of 1T'-MoTe$_2$ (see Supplemental Material).

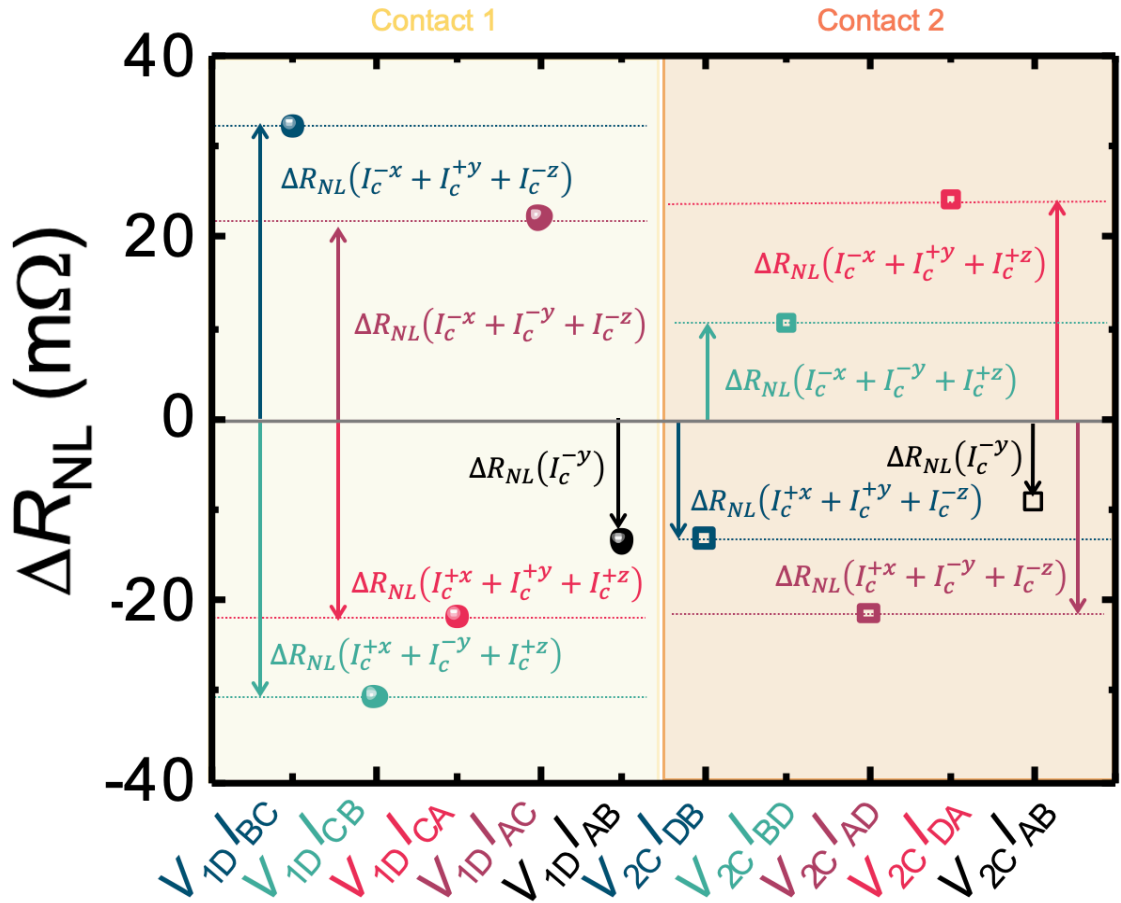

**Fig. 4.** Amplitude of the SCI signals ($\Delta R_{NL}$) for each of the eight different current paths in the 3D-current configurations as well as for the standard configuration. Each arrow is labeled with the charge current involved in the SCI process for each measurement. Contact 1 was used as a detector for the five configurations on the left side (light yellow background), while contact 2 was the detector for the five combinations on the right side (light orange background). The error bars are smaller than the dots.

By comparing the amplitudes and signs of the different SCI components in our device and assuming the three processes to occur via SHE in MoTe$_2$, we conclude that

$\sigma^y_{zx}$ presents the same sign as $\sigma^y_{zz}$ and opposite to $\sigma^y_{yz}$, and that $\sigma^y_{zx}$ is larger in magnitude than $\sigma^y_{zz}$ and $\sigma^y_{yz}$. This hierarchy of components can be compared with that obtained from DFT calculations of the spin Hall conductivity (SHC) tensor, which is one of the effects contributing to SCI. These calculations output zero values for tensor components not allowed by symmetry, such as $\sigma^y_{yz}$, while all components allowed by symmetry are non-zero (see Supplemental Material section S1). Our DFT calculations yield a finite value for the non-orthogonal SHC component $\sigma^y_{zz}$, which is symmetry allowed but has been predicted to be negligible in previous works [64]. Our calculations show that $\sigma^y_{zz}$ is not negligible but does remain smaller than the other dominant components around Fermi energy such as $\sigma^y_{zx}$. Our experimental observation that $|\sigma^y_{zx}| \gg |\sigma^y_{zz}|$ would therefore be consistent with a spin Hall contribution to these effects. Our calculations however predict opposite signs for these two components. Given the small value of $\sigma^y_{zz}$ a sign change is not implausible due to contributions from extrinsic effects such as skew-scattering and/or EE SCI mechanisms, which can never be excluded from our observations and preclude quantitative comparisons with intrinsic spin Hall effect calculations

## IV.   CONCLUSIONS

In conclusion, we have performed SCI experiments in a graphene/MoTe$_2$ van der Waals heterostructure-based LSV, both in the standard and the 3D-current configurations. Our measurement configurations allow access to three different SCI components enabled by the low resistance of the graphene/MoTe$_2$ interface. First, we confirm an unconventional SCI process where a charge current applied along *y* generates a spin polarization parallel to it, which can neither be explained by the SHE in MoTe$_2$ nor by the EE in the proximitized graphene without further lowering the system symmetry by shear strain or twisting between the layers. Also in the same device, we detected an orthogonal SCI component that has never been experimentally observed before using LSVs, where the charge current applied along *x* generates a spin polarization along *y*. As both the SHE in MoTe$_2$ and EE in the proximitized graphene allow this SCI configuration, either or both of them could be its plausible origins. Finally, we observed a third component of SCI where the charge current along *z*, generates spins polarized along *y* parallel to the spin current direction. This non-orthogonal component is allowed by the crystal symmetry of 1T'-MoTe$_2$ and can only be attributed to SHE in MoTe$_2$. To further confirm its origin, we have performed DFT calculations showing that this component is not only allowed by symmetry but the intrinsic $\sigma^y_{zz}$ is also non-zero. Considering all the above, a charge current in any direction within graphene/MoTe$_2$ heterostructure will generate spins polarized along the long axis of the MoTe$_2$, which corresponds to the direction along the Mo-atom chain. Furthermore, when contributions from all three charge current directions add-up, the spin polarization signal doubles the best value obtained for current flowing in only one direction. The rich spin-charge interconversion configurations in this system, with remarkable amplitude signals at room temperature, add flexibility in the architecture of next generation spin-based devices.

**Acknowledgements**

This work is supported by the Spanish MICINN under project RTI2018-094861-B-I00, PGC2018-101988-B-C21, PID2019-109905GB-C21, MAT2017-88377-C2-2-R and the Maria de Maeztu Units of Excellence Programme (MDM-2016-0618 and CEX2020-001038-M), by the "Valleytronics" Intel Science Technology Center, by the Gipuzkoa Regional Council under projects 2021-CIEN-000037-01 and 2021-CIEN-000070-01, and by the European Union H2020 under the Marie Sklodowska-Curie Actions (0766025-QuESTech and 794982-2DSTOP). N.O. thanks the Spanish MICINN for a Ph.D. fellowship (Grant no. BES-2017-07963). J.I.-A. acknowledges postdoctoral fellowship support from the "Juan de la Cierva - Formación" program by the Spanish MICINN (Grant No. FJC2018-038688-I). R.C. acknowledges funding from Generalitat Valenciana through grant CIDEGENT/2018/004 M.G.V. thanks support from the Deutsche Forschungs-gemeinschaft (DFG, German Research Foundation) GA3314/1-1 – FOR 5249 (QUAST).

## Supplemental Material for

*Unconventional charge-to-spin conversions in graphene/MoTe₂ van der Waals heterostructures*

Nerea Ontoso[1], C.K. Safeer[1], Franz Herling[1], Josep Ingla-Aynés[1], Haozhe Yang[1], Zhendong Chi[1], Iñigo Robredo[2], Maia G. Vergniory[2,3], Fernando de Juan[3,4], M. Reyes Calvo[5,6], Luis E. Hueso[1,4], Fèlix Casanova[1,4]

[1] CIC nanoGUNE BRTA, 20018 Donostia-San Sebastian, Basque Country, Spain.
[2] Max Planck Institute for Chemical Physics of Solids, Dresden D-01187, Germany
[3] Donostia International Physics Center, 20018 Donostia-San Sebastian, Basque Country, Spain
[4] IKERBASQUE, Basque Foundation for Science, 48009 Bilbao, Basque Country, Spain.
[5] Departamento de Física Aplicada, Universidad de Alicante, 03690 Alicante, Spain.
[6] Instituto Universitario de Materiales de Alicante (IUMA), Universidad de Alicante, 03690 Alicante, Spain.

### S1. Spin Hall Conductivity DFT calculations

To compute the spin Hall conductivity (SHC) tensor components, we first performed density functional theory (DFT) calculations as implemented in the Vienna ab initio simulation package (VASP)[1-4]. The interaction between ion cores and valence electrons was treated by the projector augmented-wave method[5], the generalized gradient approximation (GGA) for the exchange-correlation potential with the Perdew-Burke-Ernkzerhof for solids parameterization[6] and spin-orbit coupling was taken into account by the second variation method[7]. A Monkhorst-Pack centered at $\Gamma$ k-point grid of (15x7x5) for reciprocal space integration and 500 eV energy cutoff of the plane-wave expansion have been used. We constructed maximally localized Wannier functions[8] and performed SHC tensor components via Wannier interpolation calculation as implemented in WannierBerri[9]. Our calculations output non-zero values of the SHC components allowed by symmetry:

$$\sigma_{ij}^x = \begin{pmatrix} 0 & \sigma_{xy}^x & 0 \\ \sigma_{yx}^x & 0 & \sigma_{yz}^x \\ 0 & \sigma_{zy}^x & 0 \end{pmatrix} \quad \sigma_{ij}^y = \begin{pmatrix} \sigma_{xx}^y & 0 & \sigma_{xz}^y \\ 0 & \sigma_{yy}^y & 0 \\ \sigma_{zx}^y & 0 & \sigma_{zz}^y \end{pmatrix} \quad \sigma_{ij}^z = \begin{pmatrix} 0 & \sigma_{xy}^z & 0 \\ \sigma_{yx}^z & 0 & \sigma_{yz}^z \\ 0 & \sigma_{zy}^z & 0 \end{pmatrix}$$

In Fig. S1, we show the computed non-zero SHC components for an energy window of [-0.6,0.6] eV with 1000 subdivisions. The two components experimentally detected in this work are highlighted with a thicker line in Fig. S1b. We observe that all allowed components show significant dependence on the Fermi level and in particular several sign changes, oscillating within an order of magnitude of 200 $[\hbar/e]\,S/cm$. Our results in Fig. S1 are similar in magnitude to those reported in Ref. 10 and are consistent with the largest components being $\sigma_{yx}^z$ and $\sigma_{xy}^z$ which have opposite signs. Other components show a larger discrepancy and, in particular, we do not find $\sigma_{zz}^y$ to vanish to the precision shown in Ref.[10]. A possible source of these differences, beyond the possible Fermi level mismatch, is that in Ref.[10] the effective tight binding Hamiltonian is built from a basis of atomic orbitals, while we construct maximally localized Wannier functions. Our calculations therefore show that $\sigma_{zz}^{y}$ can indeed have a spin Hall origin.

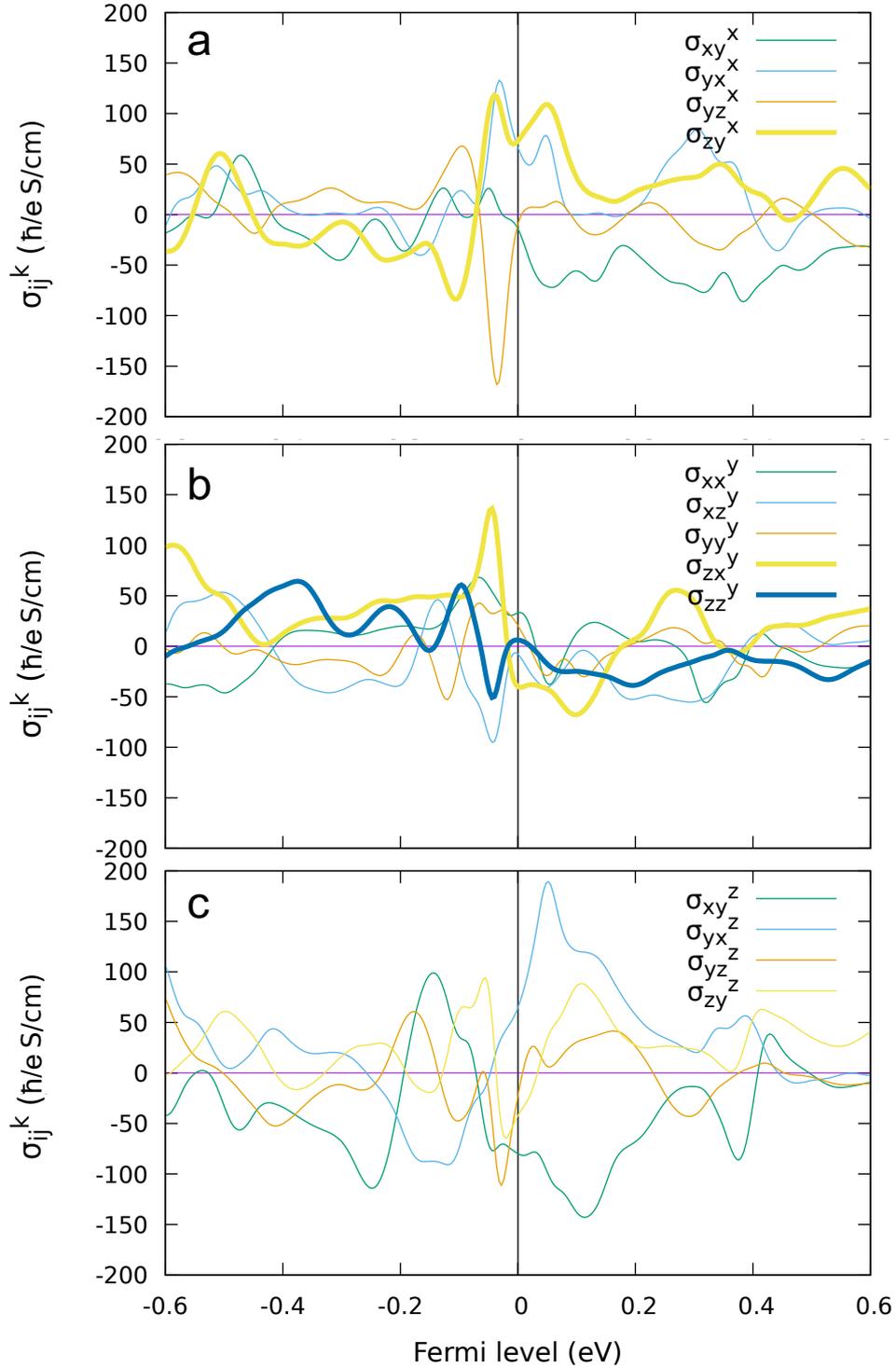

**Figure S1**: SHC tensor component calculation as a function of the chemical potential for all non-zero components for spin polarization in the (a) x, (b) y and (c) z direction. Bold lines represent the experimentally measured components.

## S2. Polarized Raman characterization of MoTe$_2$ bulk and flakes

Refs.[11,12] suggest that, while bulk crystals of MoTe$_2$ undergo a structural phase transition from 1T' to 1T$_d$ phase, thin exfoliated flakes do not show any signature of transitioning and might stay on the 1T' phase even at low temperatures. Our measurements of resistance versus temperature in MoTe$_2$ flakes, as the one shown in Fig. 1c in the main text, show indeed no signatures of a structural phase transition, i.e., no sudden changes of resistance occur when sweeping between 4 and 300 K, nor any hysteresis between the cooling and heating sweeps.

Furthermore, we performed a polarized Raman characterization of bulk and exfoliated flakes of MoTe$_2$ at variable temperature. If a transition from 1T' to 1T$_d$ phase occurs, polarized Raman spectroscopy should feature a splitting of the ~133 cm$^{-1}$ Raman mode[12,13].

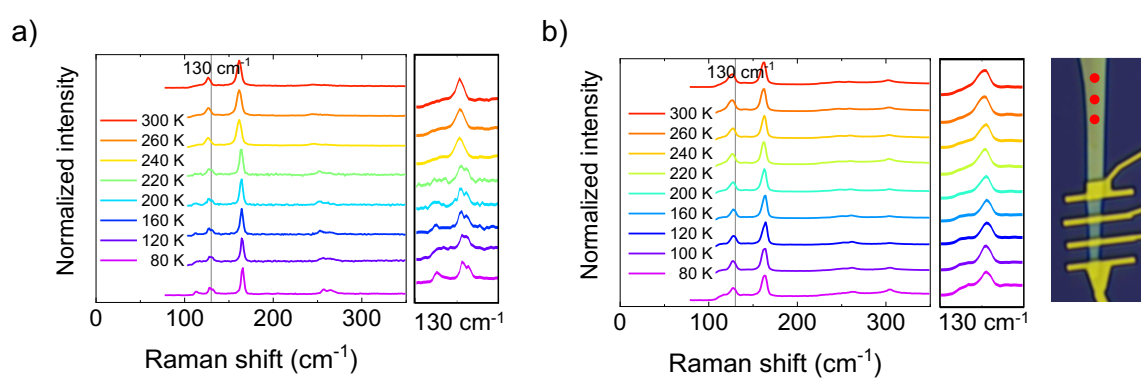

**Figure S2**. Polarized Raman spectroscopy as a function of temperature for MoTe$_2$ (a) bulk crystal and (b) exfoliated elongated flake of ~50 nm thickness – a microscope image is shown in the right panel. Raman spectra are recorded for parallel polarization in excitation and detection aligned along the Mo-chain direction of MoTe$_2$. In (b) Raman data for each temperature is the average of measurements taken at the red circles on the MoTe$_2$ flake.

Figure S2a shows polarized Raman spectroscopy measurements performed with green laser (532 nm wavelength) excitation for temperatures ranging from 300 K to 80 K in a bulk MoTe$_2$ crystal from which our flakes are exfoliated. A splitting of the Raman peak at ~133 cm$^{-1}$ appears for temperatures below 240 K indicating a transition from 1T' to the 1T$_d$ phase. In contrast, Fig. S2b shows Raman spectra acquired on a flake exfoliated from the bulk piece. No splitting of the 133 cm$^{-1}$ Raman peak can be observed, suggesting that our flakes stabilized in the 1T' phase, at least down to 80 K.

Polarized Raman spectroscopy, together with temperature-dependent resistance measurements of the 1T'-MoTe$_2$ flakes, suggest that exfoliated flakes stabilize in the 1T' phase. These measurements are in agreement with other reports where no phase transition is observed in 1T'-MoTe$_2$ flakes down to 5 K[11].

## S3. Hanle spin precession measurements

To detect Hanle spin precession signals, SCC measurements both in the standard and 3D configuration were performed with sweeping the magnetic field along the *x*-axis, the in-plane hard axis for the Co electrodes (Fig. S3a-c and Fig. S4a-c). Firstly, an "S-shape" signal background is observed, which might correspond to the detection of an x-polarized spin current as the Co electrodes align their magnetization to the x-axis at high enough $B_x$ field[14]. This may originate from conventional spin Hall effect in the MoTe$_2$ or Rashba-Edelstein in the proximitized graphene. However, we have recently demonstrated that this type of signals can also arise just from ordinary Hall effect in graphene induced by stray fields[15] and one need to perform Hanle precession around $B_z$ to confirm its spin origin, as we did in Ref.[14]. Independently of its origin, the *S-shape* signals overlap either if the measurement starts with the detector magnetization pointing in the +y ($R_{NL}^\uparrow$) or in the -y ($R_{NL}^\downarrow$) direction.

On the contrary, for low values of $B_x$, opposite sign contributions to the non-local resistance appear for $R_{NL}^\uparrow$ and $R_{NL}^\downarrow$ measurements (Fig. S3b,c and Fig. S4b,c). Those signals arise from the precession of *y*-polarized spins as they travel through the graphene channel due to the applied $B_x$ field. Although the curves are incomplete due to the switching of the Co electrodes, the subtraction signal ($\Delta R_{NL} = R_{NL}^\uparrow - R_{NL}^\downarrow$) removes the antisymmetric *S-shape* background and results in symmetric signals (at a field range below the unavoidable switching fields) arising from detection of precessing spins. Hanle spin precession measurements demonstrate the spin origin of the non-local signals, both for the standard (Fig. S3d,e) and the 3D (Fig. S4d,e) measurement configurations.

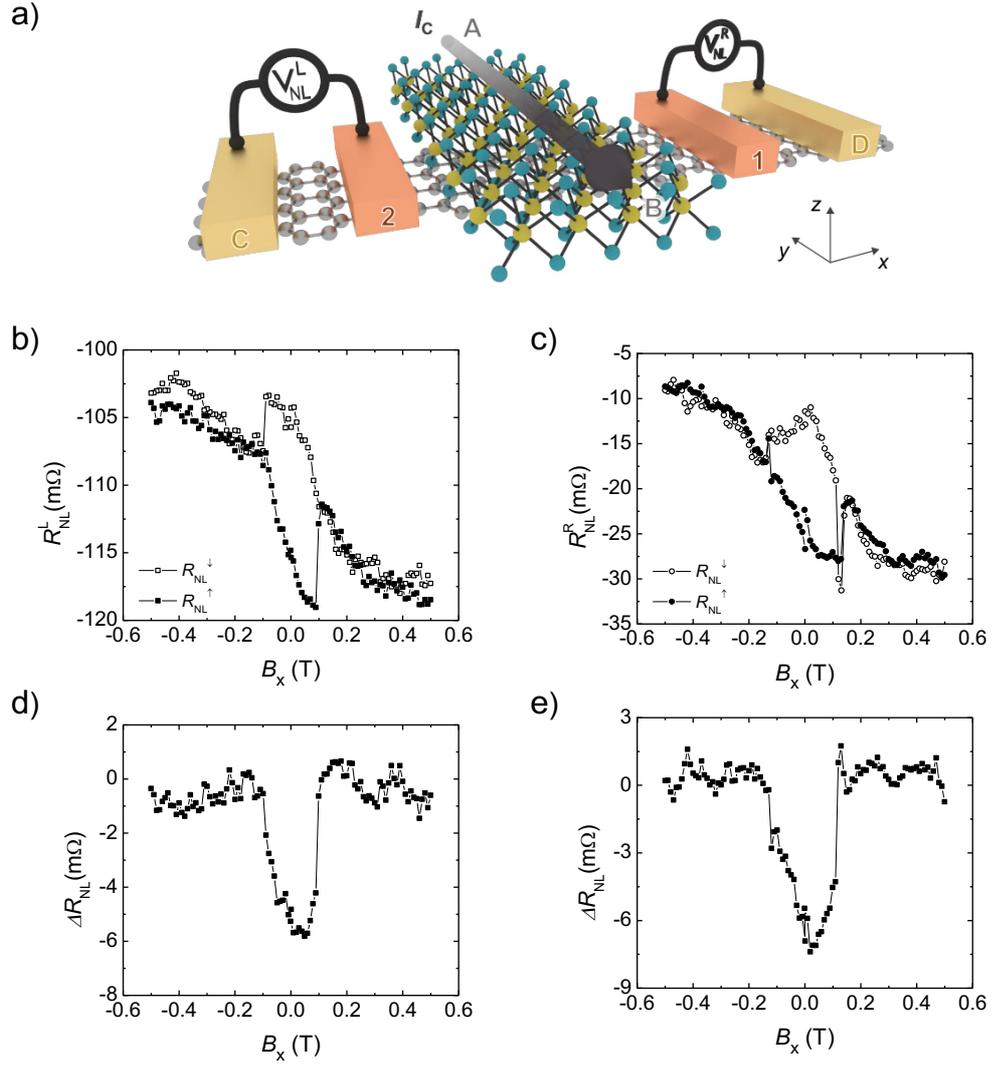

**Figure S3**. (a) Sketch of the standard SCI measurement configuration. (b,c) Non-local resistance measured while sweeping the magnetic field along the FM in-plane hard axis ($B_x$) (b) using the left Co electrode 2 as a detector and (c) using the right Co electrode 1. $R_{NL}^{\uparrow}$ and $R_{NL}^{\downarrow}$ indicate measurements performed from Co electrode magnetization set in the easy axis (*y*) positive and negative directions. (d,e) Hanle precession signals calculated as $\Delta R_{NL} = R_{NL}^{\uparrow} - R_{NL}^{\downarrow}$ using data in panels (b) and (c), respectively.

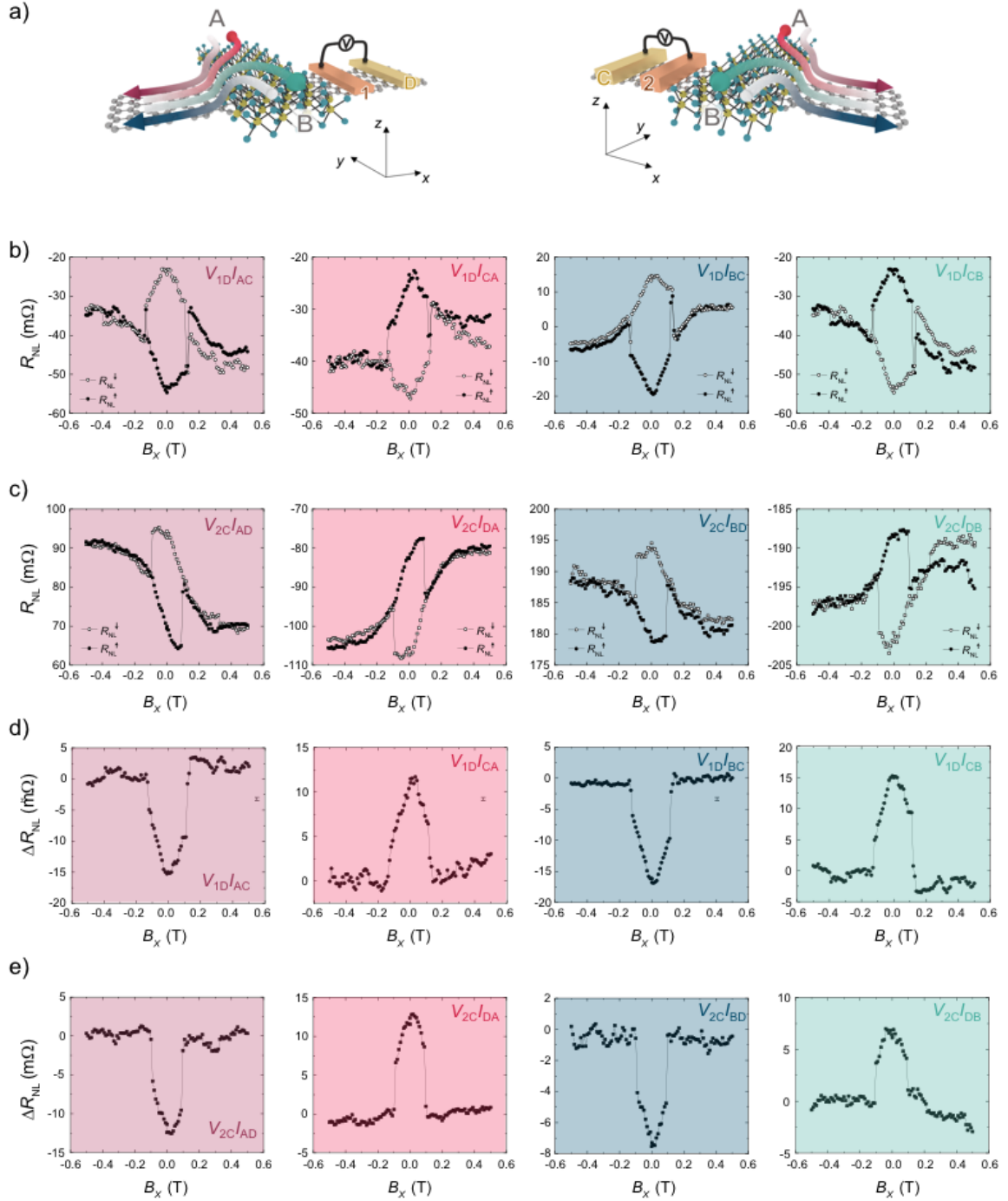

**Figure S4**. (a) Sketch of all the 3D-current SCI measurement configurations using contact 1 (left panel) and contact 2 (right panel) as FM detector. (b,c) Non-local resistance measured using contact 1 (b) and contact 2 (c) as FM detector while sweeping the magnetic field along the FM in-plane hard axis ($B_x$) for the different current paths indicated by the arrows in panels (a). $R_{NL}^{\uparrow}$ and $R_{NL}^{\downarrow}$ indicate measurements performed from Co electrode magnetization set in the easy axis ($y$) positive and negative directions. (d,e) Hanle precession signals calculated as $\Delta R_{NL} = R_{NL}^{\uparrow} - R_{NL}^{\downarrow}$ using data in panels (b) and (c), respectively.